\documentclass[prd,preprint,superscriptaddress,showpacs,nofootinbib,%
tightenlines ]{revtex4}
\usepackage{epsfig}
\usepackage{amssymb}
\newcommand{\gA}{\stackrel{\circ}{g}_{\!A}}
\newcommand{\m}{\,\stackrel{\circ}{m}\,}

\newcommand{\ben}{\begin{displaymath}}
\newcommand{\een}{\end{displaymath}}
\newcommand{\be}{\begin{equation}}
\newcommand{\ee}{\end{equation}}
\newcommand{\bea}{\begin{eqnarray}}
\newcommand{\eea}{\end{eqnarray}}
\begin{document}
\preprint{MKPH-T-05-04}
\title{Including the $\Delta(1232)$ resonance in baryon chiral perturbation
theory}
\author{C.~Hacker}
\affiliation{Institut f\"ur Kernphysik, Johannes
Gutenberg-Universit\"at, D-55099 Mainz,
Germany}
\author{N.~Wies}
\affiliation{Institut f\"ur Kernphysik, Johannes
Gutenberg-Universit\"at, D-55099 Mainz,
Germany}
\author{J.~Gegelia}
\affiliation{Institut f\"ur Kernphysik, Johannes
Gutenberg-Universit\"at,  D-55099 Mainz,
Germany}
\affiliation{
High Energy Physics Institute, Tbilisi State University,
Tbilisi, Georgia}
\author{S.~Scherer}
\affiliation{Institut f\"ur Kernphysik, Johannes
Gutenberg-Universit\"at, D-55099 Mainz,
Germany}
\date{May 6, 2005}

\begin{abstract}

   Baryon chiral perturbation theory with explicit $\Delta(1232)$ degrees
of freedom is considered.
   Using the extended on-mass-shell renormalization scheme, a manifestly
Lorentz-invariant effective field theory with a systematic power counting
is obtained.
   As applications we discuss the mass of the nucleon, the pion-nucleon
sigma term, and the pole of the $\Delta$ propagator.
\end{abstract}

\maketitle

\section{Introduction}

   Chiral perturbation theory (ChPT) has been very successful in describing
the vacuum sector of QCD at low energies
\cite{Weinberg:1979kz,Gasser:1984yg,Gasser:1984gg}.
   The sector with one baryon is more complicated.
   There, problems of obtaining a systematic power counting were
encountered in the manifestly Lorentz-invariant formulation
of the corresponding effective field theory (EFT) \cite{Gasser:1988rb}.
   These problems have been handled in the framework of the so-called
heavy-baryon chiral perturbation theory (HBChPT)
\cite{Jenkins:1991jv,Bernard:1992qa}.
   Later the systematic power counting has also been restored in the original
manifestly Lorentz-invariant formulation
\cite{Tang:1996ca,Ellis:1998kc,Becher:1999he,
Gegelia:1999gf,Gegelia:1999qt,Fuchs:2003qc}.

   The strong coupling of the $\Delta(1232)$ resonance to the $\pi N$ channel
and the relatively small mass difference between the nucleon and the delta
motivate the inclusion of the $\Delta$ as an explicit dynamical degree of
freedom in baryon chiral perturbation theory (BChPT).
   This has been done in a systematic way in the heavy-baryon formulation
using the so-called ``small scale expansion'' (see, e.g.,
Ref.~\cite{Hemmert:1997ye} for an overview).
   A different power counting for the EFT with an explicit $\Delta$
degree of freedom has been considered in
Refs.~\cite{Pascalutsa:2002pi,Pascalutsa:2004je}.

   When explicitly including the $\Delta$ resonance in BChPT,
one has to deal with the whole complexity of
a {\it consistent interaction of higher-spin fields}
(see, e.g., Refs.~\cite{Dirac:1936tg,Fierz:1939ix,Johnson:1960vt,Velo:1969bt}).
   The problem is that, in a Lorentz-invariant formulation of a field
theory involving particles of higher spin ($s\geq 1$), one necessarily
introduces unphysical degrees of freedom.
   Physical degrees of freedom are defined on a surface specified by
constraints.
   It turns out that it is rather non-trivial to write
down interaction terms which respect the constraint structure of
the theory.
   Various suggestions for constructing consistent
interactions involving spin-3/2 particles can be found in, e.g.,
Refs.~\cite{Nath:1971wp,Tang:1996sq,Pascalutsa:1998pw,
Pascalutsa:1999zz,Deser:2000dz,Pascalutsa:2000kd,
Pilling:2004wk,Pilling:2004cu}.

   In this context we note that, as the low-energy EFT deals with small
fluctuations of field variables around the vacuum, the problems showing up
for stronger fields are not relevant to this theory.
   For such configurations the higher-order terms (infinite in number)
generate contributions to physical quantities which are no longer suppressed
by powers of small expansion parameters.
   Therefore, for large fluctuations the conclusions drawn from an analysis
of a finite number of terms of the effective Lagrangian cannot be trusted.
   On the other hand, for small fluctuations around the vacuum one requires
that the theory describes the right number of degrees of freedom in a
self-consistent way.
   The interaction terms can be analyzed order by order in a small parameter
expansion.
   Such an analysis leads to non-trivial constraints on the possible
interactions.

   In the present paper we consider the manifestly Lorentz-invariant
form of BChPT with explicit $\Delta$
degrees of freedom.
   When using the standard formulation in combination with dimensional
regularization one may face difficulties with respect to constructing
the correct Lagrangian for spin-3/2 particles in $n$ space-time dimensions.
   Therefore, we apply the higher-derivative formulation of
Ref.\ \cite{Djukanovic:2004px} which is explicitly defined in four
space-time dimensions.
   In order to generate a systematic power counting in the relevant
EFT, we need to choose a suitable renormalization condition
\cite{Becher:1999he,Gegelia:1999gf,Gegelia:1999qt,Fuchs:2003qc}.
   While the infrared regularization of Ref.~\cite{Becher:1999he} has
been reformulated in a form which is also applicable to the effective theory
with explicit resonance degrees of freedom \cite{Schindler:2003xv}, we
use the extended on-mass-shell (EOMS) renormalization scheme of
Ref.~\cite{Fuchs:2003qc} in this work.
   Using as examples the one-loop contributions to the nucleon and
$\Delta$ self-energies we demonstrate that there is a systematic power
counting in the manifestly Lorentz-invariant formulation of the considered
EFT.
   Other approaches, such as the twisted mass ChPT
and chiral extrapolations in the framework of the covariant small
scale expansion have recently been discussed in
Refs.\ \cite{Walker-Loud:2005bt} and \cite{Bernard:2005fy}, respectively.

\section{The effective Lagrangian}

   In this section we will briefly discuss those elements of the most
general effective Lagrangian which are relevant for the subsequent
calculation of the nucleon mass and the pole of the $\Delta$
propagator at order $p^3$.\footnote{Here, $p$ stands for small
parameters of the theory like the pion mass and the
$\Delta$-nucleon mass difference.}
   All parameters and fields correspond to renormalized quantities.
Counterterms are not explicitly shown.

\subsection{Non-resonant Lagrangian}

   The non-resonant part of the effective Lagrangian
consists of the sum of the purely mesonic and the $\pi N$
Lagrangians, respectively,
\begin{displaymath}
{\cal L}_{\rm eff}={\cal L}_{\pi}+{\cal L}_{\pi N},
\end{displaymath}
both of which are organized in a (chiral) derivative and
quark-mass expansion (see, e.g., Ref.\ \cite{Scherer:2002tk} for an
introduction),
\begin{eqnarray*}
{\cal L}_{\pi}&=&{\cal L}_2+{\cal L}_4+\cdots,\\
{\cal L}_{\pi N}&=&{\cal L}_{\pi N}^{(1)}+{\cal L}_{\pi N}^{(2)}+
{\cal L}_{\pi N}^{(3)} +\cdots,
\end{eqnarray*}
where the subscripts (superscripts) in ${\cal L}_{\pi}$ (${\cal L}_{\pi N}$)
refer to the order in the expansion.
    The lowest-order mesonic Lagrangian reads \cite{Gasser:1984yg}
\begin{equation}
\label{l2} {\cal L}_2=\frac{F^2}{4}\mbox{Tr}\left[D_\mu U \left(
D^\mu U\right)^\dagger\right] +\frac{F^2}{4}\mbox{Tr} \left( \chi
U^\dagger+ U\chi^\dagger \right).
\end{equation}
   The pion fields are contained in the unimodular unitary $(2\times 2)$
matrix $U$ which, under a
{\em local} $\mbox{SU}(2)_L\times\mbox{SU}(2)_R\times\mbox{U}(1)_V$
transformation denoted by the group element
$(V_L(x), V_R(x),\exp[-i\Theta(x)])$, transforms as
\begin{equation}
\label{utrans}
U\mapsto U'=V_R U V_L^\dagger.
\end{equation}
   Introducing external fields $l_\mu$ and $r_\mu$ transforming as
\begin{eqnarray*}
l_\mu & \mapsto & V_L l_\mu V_L^\dagger+iV_L\partial_\mu V_L^\dagger,\\
r_\mu & \mapsto & V_R r_\mu V_R^\dagger+iV_R\partial_\mu V_R^\dagger,
\end{eqnarray*}
the covariant derivative is defined as
$$
D_\mu U=\partial_\mu U-i r_\mu U+i U l_\mu.
$$
   In Eq.\ (\ref{l2}), $F$ denotes the pion-decay constant in the chiral
limit: $F_\pi=F[1+{\cal O}(\hat{m})]=92.4$ MeV.
   Moreover,
\begin{displaymath}
\chi =2 B (s+i p)
\end{displaymath}
contains external scalar and pseudoscalar fields transforming as
$\chi\mapsto V_R\, \chi\, V_L^\dagger$.
   We work in the isospin-symmetric limit $m_u=m_d=\hat{m}$
and the lowest-order expression for the squared pion mass is
$M^2=2 B \hat{m}$, where $B$ is related to the scalar quark condensate
in the chiral limit \cite{Gasser:1984yg}, $\langle 0|\bar{u} u|0\rangle
=\langle 0|\bar{d} d|0\rangle=-F^2 B$.

   To improve the ultraviolet behavior of the pion propagator and
to regulate the loop diagrams calculated in this work, we use the
higher-derivative formulation of Ref.~\cite{Djukanovic:2004px}.
   We include the following additional terms in the effective Lagrangian:
$$
{\cal L}_{\pi\pi}^{\rm reg}= \sum\limits_{n=1}^{N_\pi}
\frac{X_n}{4} \ \frac{F^2}{4} \ {\rm Tr} \left( \left\{
\left(D^2\right)^nU U^\dagger -U\left[
\left(D^2\right)^nU\right]^\dagger \right\} \left[  D^2U
U^\dagger-U\left(D^2 U\right)^\dagger -\chi U^\dagger
+U\chi^\dagger \right] \right),
$$
where $D^2U=D_\alpha D^\alpha U$ and $X_n$ are free parameters.
For our calculations we take $N_\pi =4$ and choose the parameters $X_i$ as
\begin{eqnarray}
X_1 & = & \frac{4}{\Lambda^2},\nonumber\\
X_2 & = & \frac{6}{\Lambda^4},\nonumber \\
X_3 & = & \frac{4}{\Lambda^6},\nonumber\\
X_4 & = & \frac{1}{\Lambda^8}, \label{xis}
\end{eqnarray}
resulting in the modified pion Feynman propagator
\begin{equation}
\Delta^\Lambda_\pi(p)= \frac{1}{p^2-M^2+i 0^+} \,
\frac{\left(-\Lambda^2\right)^4}{\left( p^2-\Lambda^2+i
0^+\right)^4}. \label{regpprop}
\end{equation}
Here, $\Lambda$ is a free parameter which plays the role of a
regulator in the loop diagrams.

   Let $\Psi=(p,n)^T$ denote the nucleon field with two four-component
Dirac fields $p$ and $n$ describing the proton and neutron, respectively,
transforming as
\begin{equation}
\label{psitrans}
\Psi\mapsto\Psi'= \exp[-i\Theta]K[V_L,V_R,U]\Psi,
\end{equation}
where
\begin{equation}
\label{Kdef}
K[V_L,V_R,U]=\sqrt{V_R U V_L^\dagger}^{-1} V_R \sqrt{U}.
\end{equation}
   The $\pi N$ Lagrangian is bilinear in $\bar\Psi$ and $\Psi$ and
involves the quantities $u$, $u_\mu$, $\Gamma_\mu$, and $\chi_{\pm}$
(and their derivatives):
\begin{eqnarray*}
u^2&=&U,\\
u_\mu&=&i\left[ u^{\dagger}\left(\partial_\mu-i
r_\mu\right)u-u \left( \partial_\mu-i l_\mu \right)
u^{\dagger}\right],\\
\Gamma_\mu&=&\frac{1}{2}\, \left[ u^{\dagger}\left(\partial_\mu-i
r_\mu\right)u + u \left(
\partial_\mu-i l_\mu\right) u^{\dagger}\right],\\
\chi_{\pm}&=&u^{\dagger}\chi u^{\dagger}\pm u\chi^\dagger u.
\end{eqnarray*}
   In terms of these building blocks the lowest-order Lagrangian reads
\cite{Gasser:1988rb}
\begin{equation}
{\cal L}_{\pi N}^{(1)}=\bar \Psi \left( i\gamma_\mu D^\mu -m
+\frac{1}{2} \stackrel{\circ}{g}_A\gamma_\mu \gamma_5 u^\mu\right) \Psi,
\label{lolagr}
\end{equation}
where the covariant derivative is defined as
$$D_\mu\Psi = (\partial_\mu +\Gamma_\mu-i v_\mu^{(s)})\Psi $$
with the external field transforming as $v_\mu^{(s)}\mapsto v_\mu^{(s)}-\partial_\mu\Theta$.
   Finally, $m$ denotes the mass of the nucleon at leading order in
the expansion in small parameters and
$\gA$ refers to the chiral limit of
the axial-vector coupling constant.

   For our purposes, we only need to consider one of the seven structures
of the Lagrangian at ${\cal O}(p^2)$ \cite{Gasser:1988rb}
\begin{equation}
{\cal L}_{\pi N}^{(2)}=\tilde c_1 \mbox{Tr}(\chi_{+})\bar\Psi\Psi
+\cdots, \label{p2olagr}
\end{equation}
where $\tilde c_1$ refers to the coupling constant in the theory
explicitly including delta degrees of freedom.
   The Lagrangian ${\cal L}^{(3)}_{\pi N}$ does not contribute in our
calculations.

\subsection{Lagrangian of the $\Delta(1232)$ resonance}
   In order to write down the Lagrangian of the $\Delta(1232)$ resonance
[$I(J^P)=\frac{3}{2}(\frac{3}{2}^+)$] we introduce the vector-spinor
isovector-isospinor fields
\begin{equation}
\label{deltadef}
\Psi_{\mu,i}=\left(\begin{array}{c}\Psi_{\mu,i,\frac{1}{2}}\\
\Psi_{\mu,i,-\frac{1}{2}}\end{array}\right),\quad \mu=0,1,2,3,\quad i=1,2,3,
\end{equation}
i.e., for any values of $\mu$ and $i$ the component
$\Psi_{\mu,i}$ consists of an isospin doublet of Dirac spinors.
   The delta field components transform
as \cite{Tang:1996sq}
\begin{equation}
\label{deltatrans}
\Psi_{\mu,i,\alpha}\mapsto \Psi_{\mu,i,\alpha}'
=\exp[-i\Theta]{\cal K}_{ij,\alpha\beta}\Psi_{\mu,j,\beta},
\end{equation}
where
\begin{equation}
\label{calKdef}
{\cal K}_{ij,\alpha\beta}=\frac{1}{2}\mbox{Tr}(\tau_i K \tau_j K^\dagger)
K_{\alpha\beta}
\end{equation}
with $K$ defined in Eq.\ (\ref{Kdef}).
   The corresponding covariant derivative is given by
\begin{eqnarray*}
(D_\mu \Psi)_{\nu,i,\alpha}&\equiv& {\cal D}_{\mu,ij,\alpha\beta}
\Psi_{\nu,j,\beta},\\
{\cal D}_{\mu,ij,\alpha\beta}&=&
\partial_\mu\delta_{ij}\delta_{\alpha\beta}
-2i\epsilon_{ijk}\Gamma_{\mu,k}\delta_{\alpha\beta}
+\delta_{ij}\Gamma_{\mu,\alpha\beta}
-iv_\mu^{(s)}\delta_{ij}\delta_{\alpha\beta},
\end{eqnarray*}
where we parameterized $\Gamma_\mu=\tau_k\Gamma_{\mu,k}$.
   The description of Eq.\ (\ref{deltadef}) involves 6 (uncoupled) isospin
components whereas the physical delta consists of an isospin quadruplet.
   Introducing the isospin projection operators\footnote{Note that the
isovector components refer to a Cartesian isospin basis.}
\begin{eqnarray}
\xi^\frac{3}{2}_{ij,\alpha\beta}&=&\delta_{ij}\delta_{\alpha\beta}
-\frac{1}{3}(\tau_i\tau_j)_{\alpha\beta},\label{xi32}\\
\xi^\frac{1}{2}_{ij,\alpha\beta}&=&\frac{1}{3}(\tau_i\tau_j)_{\alpha\beta},
\label{xi12}
\end{eqnarray}
the leading-order Lagrangian is given by \cite{Hemmert:1997ye}\footnote{
We have explicitly included the projection operator in the definition
of the Lagrangian.
}
\begin{equation}
{\cal L}_{\Delta}^{(1)}=\bar\Psi_\mu\xi^{\frac{3}{2}}
\Lambda^{\mu\nu}\xi^{\frac{3}{2}} \Psi_\nu, \label{deltaL}
\end{equation}
where
\begin{eqnarray}
\label{Ldeltafree} \Lambda_{\mu\nu} &=& -\Biggl[(i
{D\hspace{-.70em}/\hspace{.1em}}-m_{\Delta})\,g_{\mu\nu} +i
A\,(\gamma_{\mu}D_{\nu}+\gamma_{\nu}D_{\mu})\nonumber \\
&&+\frac{i}{2}\,(3A^2+2A+1)\,\gamma_{\mu}
{D\hspace{-.70em}/\hspace{.1em}}\gamma_{\nu} + m_{\Delta}
\,(3A^2+3A+1)\,\gamma_{\mu}\gamma_{\nu}\nonumber
\\
&&+\frac{g_1}{2}\, u\hspace{-.5em}/\,\gamma_5\,
g_{\mu\nu}+\frac{g_2}{2}\,\left( \gamma_\mu u_\nu + u_\mu
\gamma_\nu\right)\gamma_5 +\frac{g_3}{2}\,\gamma_\mu u\hspace{-.5em}/\,
\gamma_5\gamma_\nu \Biggr].
\end{eqnarray}
   Here, $A$ is an arbitrary real parameter except that $A\neq -1/2$
and $m_\Delta $ denotes the mass of the $\Delta$ at leading order
in the expansion in small parameters.

   The Lagrangian of Eq.~(\ref{deltaL}) describes a system with constraints.
   Using the canonical formalism (i.e.~canonical coordinates and momenta and
the corresponding Hamiltonian) we have analyzed the structure of the
constraints in analogy with Refs.~\cite{Nath:1971wp,Pascalutsa:1998pw}.
   Demanding that the above interaction terms lead to a consistent theory with
the correct number of physical degrees of freedom we
obtain after a lengthy calculation the following
relations among the coupling constants \cite{Wies:2005}:
\begin{equation}
g_2=A\,g_1\,, \quad g_3=-\frac{1+2\,A+3\,A^2}{2} \  g_1.
\label{couplingrelations}
\end{equation}
   In other words, what seem to be independent interaction terms
from the point of view of constructing the most general Lagrangian
\cite{Hemmert:1997ye},
turn out to be related once the self consistency conditions are imposed.
   This situation is similar to the case of the universal $\rho$-meson
coupling recently discussed in Ref.~\cite{Djukanovic:2004mm}.
   There, relations among coupling constants were obtained from the
requirement of the consistency of EFT with respect to renormalization.

   The Lagrangian of Eq.~(\ref{deltaL}) with the couplings of
Eq.~(\ref{couplingrelations}) is invariant under the set of
transformations
\begin{eqnarray}
\Psi_\mu &\rightarrow& \Psi_\mu + a
\gamma_\mu\gamma^\nu\Psi_\nu\,, \label{fieldtr} \\
A &\rightarrow& \frac{A-2\,a}{1+4\,a}\label{Atrans}
\end{eqnarray}
which are often referred to as a point transformation
\cite{Nath:1971wp,Tang:1996sq}.
   The change of field variables of Eq.\ (\ref{fieldtr}) generates a
(new) Lagrangian ${\cal L}(\Psi_\mu,g_1,A+2a+4a A)$ which,
as a consequence of the equivalence theorem \cite{Kamefuchi:1961sb},
must yield the same observables as the original Lagrangian
${\cal L}(\Psi_\mu,g_1,A)$.
   Since $a$ can be chosen arbitrarily except that $a\neq -1/4$, physical
quantities cannot depend on $A$ \cite{Nath:1971wp,Tang:1996sq}.
   We will use $A=-1$ in our calculations.
   It is worth emphasizing that we did not require the invariance under the
point transformation to begin with; rather it comes out
automatically as a consequence of consistency in the sense of
having the right number of degrees of freedom.
   Moreover, demanding the invariance under the point transformation alone
would not be sufficient to obtain the relations of
Eq.~(\ref{couplingrelations}).
   For example, in Ref.~\cite{Tang:1996sq} three $\pi\Delta\Delta$
interactions involving one overall coupling constant and two additional
``off-shell parameters'' were introduced so as to preserve the
invariance under the point transformation.
   It was then shown that the contributions to physical quantities, generated by
the two interaction terms corresponding to the above coupling
constants $g_2$ and $g_3$, can be systematically included in
redefinitions of coupling constants of {\it an infinite number} of
local terms in the Lagrangian.
   As we are unable to analyze all of these structures and decide if they
are allowed by consistency conditions (in the sense of generating
the correct number of degrees of freedom), we choose to use the
above values of Eq.~(\ref{couplingrelations}), which are certainly
consistent.

   The effective Lagrangian of Eq.~(\ref{deltaL}) is also invariant
under the following local transformations
\begin{equation}
\Psi_{\mu,i} (x)\to \Psi_{\mu,i}(x)+\tau_i\alpha_\mu(x),
\label{invtransf}
\end{equation}
where $\alpha_\mu$ is an arbitrary vector-spinor isospinor function.
   This is due to the fact that we use six isospin degrees of freedom
$\Psi_{\mu,i,\alpha}(x)$ instead of four physical isospin degrees of
freedom.
   We could make use of Eq.~(\ref{phfields}) and rewrite the Lagrangian in
terms of the physical fields, but for reasons of convenience we prefer to work
with the gauge-invariant Lagrangian.

   The quantization of the effective Lagrangian of Eq.~(\ref{deltaL})
with the gauge fixing condition $\tau_i \Psi_{\mu,i} =0$ leads to
the following Feynman propagator\footnote{With this choice we
associate a factor $iS^{\mu\nu}_0(p)$ with an internal delta line
of momentum $p$.}
\begin{equation}
S^{\mu\nu}_{0,ij,\alpha\beta}(p)=\xi^{\frac{3}{2}}_{ij,\alpha\beta}
S^{\mu\nu}_0(p),
\label{deltapr0}
\end{equation}
where
\begin{eqnarray*}
S^{\mu\nu}_0(p) &=&
-\frac{p \hspace{-.45em}/\hspace{.1em} +m_{\Delta}}{p^2-m_{\Delta}^2+i0^+}\,
\Biggl [ g^{\mu\nu}-\frac{1}{3}\,\gamma^\mu\gamma^\nu +\frac{1}{3
m_{\Delta}}\left( p^\mu\gamma^\nu-\gamma^\mu
p^\nu\right)-\frac{2}{3 m_{\Delta}^2}\,p^\mu p^\nu\Biggr]\nonumber \\
&&+\frac{1}{3 m_{\Delta}^2}\,\frac{1+A}{1+2 A}\,\Biggl\{ \left[
\frac{A}{1+2 A}\,m_{\Delta}-\frac{1+A}{2 (1+2 A)}
p \hspace{-.45em}/\hspace{.1em} \right] \gamma^\mu\gamma^\nu - \gamma^\mu
p^\nu
 -\frac{A}{1+2 A}\,p^\mu
\gamma^\nu \Biggr\}.
\end{eqnarray*}
   In particular, choosing $A=-1$ results in the most convenient
expression for the free delta Feynman propagator.

   From the Lagrangian at ${\cal O}(p^2)$ we only need one term,
namely,
\begin{equation}
{\cal L}_{\Delta}^{(2)}= - c_1^\Delta
\mbox{Tr}(\chi_{+})\bar\Psi_{\mu,i}\,
\xi_{ij}^{\frac{3}{2}}\,g^{\mu\nu}\, \Psi_{\nu,j}.
\label{deltaLNLO}
\end{equation}

\subsection{$\pi N \Delta$ interaction term}

   The leading-order $\pi N\Delta$ interaction Lagrangian can be
written as
\begin{equation}
{\cal L}_{\pi N\Delta}^{(1)}= - g \,\bar{\Psi}_{\mu,i}
\,\xi^{\frac{3}{2}}_{ij} \,(g^{\mu\nu}
+\tilde{z}\,\gamma^{\mu}\gamma^{\nu})\, u_{\nu,j}\, \Psi +h.c.\,,
\label{pND}
\end{equation}
where we parameterized $u_\mu=\tau_k u_{\mu,k}$\,,  and $g$ and
$\tilde z$ are coupling constants.
   The analysis of the structure of constraints yields
\begin{equation}
\tilde z= \frac{3\,A+1}{2}.
\label{zcond}
\end{equation}
   Again, the interaction term of Eq.~(\ref{pND}) with the coupling
constants $g$ and $\tilde z$ constrained by Eq.~(\ref{zcond}) is invariant
under the point transformation of Eqs.~(\ref{fieldtr}) and (\ref{Atrans}).

\subsection{Power counting}
   We organize our perturbative calculations by applying the standard
power counting of Refs.~\cite{Weinberg:1991um,Ecker:1995gg} to the
renormalized diagrams, i.e., an interaction vertex obtained from an
${\cal O}(p^n)$ Lagrangian counts as order $p^n$, a pion propagator
as order $p^{-2}$, a nucleon propagator as order $p^{-1}$,
and the integration of a loop as order $p^4$.
   In addition, we assign the order $p^{-1}$ to the
$\Delta$ propagator and the order $p^1$ to the mass difference
$\delta\equiv m_\Delta-m$.
   As will be demonstrated below, this power
counting is respected by the renormalized loop diagrams within the EOMS
renormalization scheme of Ref.~\cite{Fuchs:2003qc}.

\section{Nucleon mass}
\label{section_nucleon_mass}
   In this section we calculate the nucleon mass to order $p^3$.
To that end we consider the two-point function of the nucleon
\begin{equation}
-i\int d^4xe^{ip\cdot x}\langle 0|T[\Psi (x) \bar\Psi (0)]|0
\rangle = \frac{1}{p\hspace{-.45em}/\, - m - \Sigma (p\hspace{-.45em}/)},
\label{tpf}
\end{equation}
where $\Sigma$ is the self-energy of the nucleon.
   The nucleon mass is identified in terms of the pole of Eq.\ (\ref{tpf}) at
$p\hspace{-.45em}/\,=m_N$.

\begin{figure}
\epsfig{file=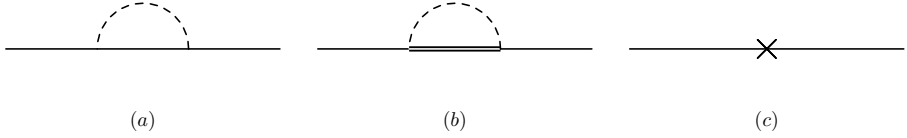, width=12truecm}
\caption[]{\label{mnloops:fig} One-loop contributions to the
nucleon self-energy at ${\cal O}(p^3)$.}
\end{figure}

   Equation (\ref{p2olagr}) generates the constant tree-level contribution
$-4\, \tilde c_1 M^2$ to the self-energy at ${\cal O}(p^2)$.
   The unrenormalized one-loop contribution 
resulting from Fig.~\ref{mnloops:fig}~(a) has the form
\begin{equation}
\label{sigmaaresult}
\Sigma_{N}^{\rm loop} = -\frac{3 \gA^2
\Lambda^8}{4 F^2}\left[ \gamma_\mu\left(
-p\hspace{-.45em}/\hspace{.1em}+m\right)\gamma_\nu \
I^{\mu\nu}_m(411) - g_{\mu\nu}\gamma_\lambda \
I^{\mu\nu\lambda}_m(411)\right],
\end{equation}
where
\begin{equation}
\left\{ I_m(abc), I^{\mu\nu}_m(abc),I^{\mu\nu\lambda}_m(abc)
\right\}=i \int \frac{d^4 k}{(2 \pi)^4} \frac{\left\{1,k^\mu
k^\nu,k^\mu k^\nu k^\lambda\right\}}{A^a B^b C^c_m},
\label{intdefIabcd}
\end{equation}
with
\begin{eqnarray}
&& A = k^2-\Lambda^2+i0^+, \nonumber \\ && B = k^2-M^2+i0^+,
\nonumber \\ && C_m = (p+k)^2-m^2+i0^+.\nonumber
\end{eqnarray}
   Substituting the results of the loop integrals of Appendix
\ref{loop_integrals} into Eq.~(\ref{sigmaaresult}), we obtain
\begin{eqnarray}
\Sigma_N^{\rm loop}(p \hspace{-.45em}/\,=m_N)&=&
-\frac{\gA^2\,m}{32\,\pi^2\,F^2}\, \left[ \Lambda^2 +7\,m^2 -
6\,m^2
\ln\left(\frac{\Lambda}{m}\right)\right]\nonumber \\
&&-\frac{\gA^2M^2}{64\,\pi^2\,F^2}\,\left[
5\,m-24\,\tilde c_1\,m^2-4\,\tilde c_1\,\Lambda^2 -
12\,m\,\ln\left(\frac{\Lambda}{m}\right)\right]\nonumber
\\
&&-\frac{3\,\gA^2M^3}{32\,\pi\,F^2}.
\label{nmsigmaa}
\end{eqnarray}
   Within the EOMS renormalization scheme \cite{Fuchs:2003qc} the first two
lines of Eq.~(\ref{nmsigmaa}) are canceled by the corresponding
contributions of the counterterm diagram of
Fig.~\ref{mnloops:fig}~(c),
leaving the last line of Eq.~(\ref{nmsigmaa}) as the
${\cal O}(p^3)$ contribution of the renormalized diagram of
Fig.~\ref{mnloops:fig}~(a).
   The unrenormalized one-loop contribution of the $\Delta$ resonance
of Fig.~\ref{mnloops:fig}~(b) reads \cite{thanks}
\begin{equation}
\label{sigmaDeltaaresult}
\Sigma_{\Delta}^{\rm loop}=
\,\frac{4\,g^2\,\Lambda^8}{3\,m_{\Delta}^2  F^2}\left[ \left(
p\hspace{-.45em}/\hspace{.1em}+m_{\Delta}\right)\left( p_\mu p_\nu
-p^2 g_{\mu\nu}\right) I^{\mu\nu}_{m_\Delta} (411) +
\gamma_\lambda \left( p_\mu p_\nu -p^2 g_{\mu\nu}\right)
I^{\mu\nu\lambda}_{m_\Delta} (411)\right].
\end{equation}
   The corresponding contribution to the mass of the nucleon is
obtained from
\begin{eqnarray}
\Sigma_{\Delta}^{\rm loop}(p \hspace{-.45em}/\,=m_N)
&=&\frac{5\,g^2}{1152\,\pi^2\,F^2}
\left[ -8\, m\,\,\Lambda^2-21\,m^3+24\,m^3\,
\ln\left(\frac{\Lambda}{m}\right)\right]\nonumber \\
&&+ \frac{g^2\,\delta}{576\,\pi^2\,F^2}
\left[28\,\Lambda^2-135\,m^2+120\,m^2\,
\ln\left(\frac{\Lambda}{m}\right)\right]\nonumber\\
&&+ \frac{g^2}{384\,\pi^2\,F^2\,m}
\left[4\,\tilde c_1\,M^2\,m\,\left( 32\,\Lambda^2+55\,m^2 \right)
- 3\,\delta^2\,\left( 8\,\Lambda^2+35\,m^2\right)\right]\nonumber\\
&&- \frac{5\,m\,g^2}{48\,\pi^2\,F^2}
\left[ \delta^2+2\,M^2\,\left( 2\,\tilde c_1\,m-1\right)\right]\,
\ln\left(\frac{\Lambda}{m}\right)\nonumber\\
&&+\frac{g^2\,\Lambda^2}{144\,\pi^2\,F^2\,m^2}
\left(11\,\delta^3 - 72\,\tilde c_1\,\delta\,M^2\,m\right)\nonumber \\
&&+ \frac{g^2}{24\,\pi^2\,F^2}\,\left[ 3\,\delta^3
-M^2\,\delta\left(7+60\,\tilde c_1\,m\right)\right]\,
\ln\left(\frac{\Lambda}{m}\right)\nonumber\\
&&+\frac{g^2}{288\,\pi^2\,F^2}
\Biggl[ 35\,\delta^3 + 6\,\delta\,M^2 \left(
1 + 25\,\tilde c_1\,m\right)\nonumber \\
&&+ 96\, \left( \delta^2-M^2\right)^{3/2}\,
\ln\left(\frac{\delta - \sqrt{\delta^2-M^2}}{M}\right)\Bigg]\nonumber \\
&& -\frac{g^2}{6\,\pi^2\,F^2}\,\left( 2\,\delta^3-3
\,M^2\,\delta\right)\,\ln\left(\frac{M}{m}\right).
\label{mndcontr}
\end{eqnarray}
   Again, in the EOMS renormalization scheme the first six lines of
Eq.~(\ref{mndcontr}) are canceled by the corresponding contributions
of the counterterm diagram of Fig.~\ref{mnloops:fig}~(c),
leaving the last three lines of Eq.~(\ref{mndcontr}) as the
delta loop contribution to the nucleon mass at ${\cal O}(p^3)$.
   Combining the tree-level result at ${\cal O}(p^2)$ with the
${\cal O}(p^3)$ one-loop contributions we obtain the following
expression for the nucleon mass:
\begin{eqnarray}
m_N&=&m-4\,\tilde c_1 M^2-{3\,\gA^2\,M^3\over
32\,\pi\,F^2} + \frac{g^2}{288\,\pi^2\,F^2} \Biggl[
35\,\delta^3 + 6\,\delta\,M^2\,\left(1
 + 25\,\tilde c_1\,m\right)\nonumber \\
&&+ 96\, \left( \delta^2-M^2\right)^{3/2}\,\ln\left(\frac{\delta
- \sqrt{\delta^2-M^2}}{M}\right)\Bigg]\nonumber \\
&&- \frac{g^2}{6\,\pi^2\,F^2}\,\left( 2\,\delta^3-3
\,M^2\,\delta\right)\,\ln\left(\frac{M}{m}\right)+{\cal O}\left( p^4\right),
\label{mass}
\end{eqnarray}
which is in agreement with Ref.~\cite{Bernard:2003xf}.

   By explicitly including the spin-3/2 degrees of freedom, terms of
higher order in the {\em chiral} expansion have been re-summed.
   In order to obtain the numerical value of these terms,
we expand Eq.~(\ref{mass}) in powers of $M$
and match the terms of orders $M^0$ and $M^2$, respectively,
to the corresponding quantities of the EFT without
explicit spin-3/2 degrees of freedom.
    Taking into account that there are no
tree-level $\Delta$ contributions to $c_1$ \cite{Becher:1999he}, we
obtain
\begin{equation}
\label{mchi}
\m=m+\frac{g^2\, \delta ^3}{3\,\pi^2\,F^2}
\ln\left(\frac{m}{2 \delta}\right)
+\frac{35\,g^2\, \delta ^3}{288\,\pi^2\, F^2 }\,,
\end{equation}
\begin{equation}
\label{cchi} c_{1}=\tilde c_1
-(1+5\,\tilde c_1\, m)\frac{5 \,g^2\, \delta}{192\,\pi^2\,F^2}
+\frac{g^2\,\delta}{8\,\pi^2\,F^2}\,\ln\left(\frac{m}{2\delta}\right),
\end{equation}
where $\m$ denotes the nucleon mass in the chiral limit and
$c_{1}$ replaces the coupling constant of Eq.\ (\ref{p2olagr}) in the
theory without spin-3/2 degrees of freedom.
   Using Eqs.~(\ref{mchi}) and (\ref{cchi}), the
nucleon mass of Eq.~(\ref{mass}) can be rewritten as
\begin{eqnarray}
m_N&=&\m-4\,c_1 M^2-{3\,\gA^2\,M^3\over 32\,\pi\,F^2} +\tilde m_N,
\label{masstilde}
\end{eqnarray}
where $\tilde m_N\sim M^4$ and contains an infinite number of
terms if expanded in powers of $M/\delta$.

    In order to calculate the numerical value of $\tilde m_N$, we make use of
$g=1.127$ as obtained from a fit to the $\Delta\to\pi N$ decay width, and
take the numerical values
\begin{eqnarray}
\label{numericalvalues} &&g_A=1.267,\quad
F_\pi=92.4\,\mbox{MeV},\quad
m_N=m_p=938.3\,\mbox{MeV},\nonumber\\
&&M_\pi=M_{\pi^+}=139.6\,\mbox{MeV},\quad
m_\Delta=1210\,\mbox{MeV},\, \delta=m_\Delta-m_N.
\end{eqnarray}
   Substituting the above values in the expression for $\tilde m_N$
results in
\begin{equation}
\tilde m_N=-5.7\, \mbox{MeV}. \label{mtilde}
\end{equation}
   We recall that an analysis of the nucleon mass up to and including
order $M^4$ \cite{Fuchs:2003kq} yields
$(882.8+74.8-15.3)\,\mbox{MeV}=942.3$ MeV for the first three terms of
Eq.\ (\ref{masstilde}).
   This analysis made use of $c_1=-0.9\,m_N^{-1}$
\cite{Becher:2001hv} as obtained from a (tree-level) fit to the
$\pi N$ scattering threshold parameters of Ref.\ \cite{Koch:bn}
and a value of 45 MeV \cite{Gasser:1990ce} for the pion-nucleon
sigma term to be discussed below.
   In other words, the explicit inclusion of the spin-3/2 degrees of
freedom does not have a significant impact on the nucleon mass.

   Applying the Hellmann-Feynman theorem \cite{Hellmann,Feynman}
to the nucleon mass
\cite{Gasser:1984yg,Gasser:1988rb,Lehnhart:2004vi}
\begin{equation}
\label{sigmaterm} \sigma=M^2 \frac{\partial m_N}{\partial M^2},
\end{equation}
the pion-nucleon sigma term to order $p^3$ reads
\begin{eqnarray}
\sigma&=&-4\,\tilde c_1 M^2-\frac{9\,\gA^2 M^3}{64\,\pi\,F^2}
+\frac{5\, g^2\,(1+ 5\,\tilde c_1\,m)\,\delta\,M^2}{48\,\pi^2 \,F^2}\nonumber\\
&&-\frac{g^2\,(\delta^2-M^2)^\frac{1}{2}\,M^2}{2\,\pi^2\,F^2}
\,\ln\left(\frac{\delta - \sqrt{\delta^2-M^2}}{M}\right)
+\frac{g^2\,\delta\,M^2}{2\,\pi^2\,F^2}\,\ln\left(\frac{M}{m}\right).
\label{sigmat}
\end{eqnarray}
   Again, expanding Eq.~(\ref{sigmat}) in powers of $M$ and using
Eq.~(\ref{cchi}), we rewrite $\sigma$ as
\begin{equation}
\sigma = -4\,c_1 M^2-\frac{9\,\gA^2M^3}{64\,\pi\,F^2}
+\tilde\sigma, \label{sigmatrewr}
\end{equation}
where $\tilde\sigma$ is of order $M^4$ and contains an infinite number of
terms if expanded in powers of $M/\delta$.
   With the numerical values of Eq.\ (\ref{numericalvalues}) we obtain from
Eq.~(\ref{sigmat})
\begin{equation}
\tilde\sigma =-10.2 \, \mbox{MeV}, \label{sigmatilde}
\end{equation}
   while the first two terms of Eq.\ (\ref{sigmatrewr}) yield
$(74.8-22.9)\,\mbox{MeV}=51.9$ MeV.
   These numbers have to be compared with the empirical values of the sigma
term extracted from data on pion-nucleon
scattering: 40 MeV \cite{Buettiker:1999ap},
$(45\pm 5)$ MeV \cite{Gasser:1990ce},
and $(64 \pm 7)$ MeV \cite{Pavan:2001wz}.
   Equation (\ref{sigmatilde}) indicates that the explicit inclusion of the
spin-3/2 degrees of freedom plays a more important role for the sigma
term than for the nucleon mass.
   However, one has to keep in mind that the sigma term only starts at order
$M^2$ and thus, on a relative scale,
is automatically more sensitive to higher-order corrections.

\section{Pole of the $\Delta$ propagator}
   Using isospin symmetry, the isospin structure of the dressed
$\Delta$ propagator is given by
\begin{equation}
S^{\mu\nu}_{ij,\alpha\beta}(p)=\xi^\frac{3}{2}_{ij,\alpha\beta}
S^{\mu\nu}(p),
\end{equation}
where $S^{\mu\nu}(p)$ is obtained by solving the equation
\begin{equation}
S^{\mu\nu}(p)=S_0^{\mu\nu}(p)
-S^{\mu\rho}(p)
\Sigma_{\rho\sigma}(p)
S_0^{\sigma\nu}(p).
\label{fullpreq2}
\end{equation}
   Here, $S_0^{\mu\nu}(p)$ refers to the free Feynman propagator of
Eq.\ (\ref{deltapr0}) and $i\,\Sigma^{\mu\nu}$ originates from the sum of
the one-particle-irreducible diagrams contributing to the two-point
function of the $\Delta$.\footnote{In analogy to the case of vector bosons,
we choose a sign convention where $i\,\Sigma^{\mu\nu}$ refers to the
components of the self-energy tensor.}
   The solution of Eq.~(\ref{fullpreq2}) has a rather complicated form
\cite{Kaloshin:2003xc}, but at this stage we are only interested in
determining the pole of the dressed propagator.
   For that purpose we may contract Eq.\ (\ref{fullpreq2})
with appropriate vector spinors,
\begin{equation}
\bar u_\mu  S^{\mu\nu} u_\nu
=\bar u_\mu S_0^{\mu\nu} u_\nu -\bar u_\mu
S^{\mu\rho}\Sigma_{\rho\sigma}S_0^{\sigma\nu} u_\nu,
\label{Dmassequation}
\end{equation}
where
\begin{eqnarray}
\gamma^\mu\, u_\mu&=&0,\nonumber\\
p^\mu\, u_\mu&=&0,
\label{ucond}
\end{eqnarray}
with corresponding expressions for the adjoints.
   We parameterize the dressed propagator and the self-energy of the $\Delta$
resonance as
\begin{equation}
\Sigma^{\mu\nu} =\sum_{a=1}^{10} \Sigma_a {\cal P}_a^{\mu\nu},
\label{deltase}
\end{equation}
\begin{equation}
S^{\mu\nu} =\sum_{a=1}^{10} S_a {\cal P}_a^{\mu\nu},
\label{deltapr}
\end{equation}
where the $\Sigma_a$ and $S_a$ are functions of $p^2$,
and the basis $\{{\cal P}_a^{\mu\nu}\}$ is specified in
Appendix \ref{decomposition}.
   Using the identities
\begin{eqnarray}
\bar u_\mu S^{\mu\nu}&=&\bar u^\nu \left(S_1+ p\hspace{-.45em}/\hspace{.1em}
S_6\right),
\label{deltaprU}\\
\bar u_\mu S_0^{\mu\nu}&=&-\frac{ \bar u^{\nu} \left(
p\hspace{-.45em}/\hspace{.1em}+m_\Delta \right)}{p^2-m_\Delta^2}\,,
\label{deltaprU01}\\
S_0^{\mu\nu} u_\nu&=&-\frac{\left(p\hspace{-.45em}/\hspace{.1em}
+m_\Delta \right) u^\mu}{p^2-m_\Delta^2},
 \label{deltaprU02}
\end{eqnarray}
we solve Eq.~(\ref{Dmassequation}) for $S_1$ and $S_6$:
\begin{eqnarray}
S_1&=&\frac{ m_\Delta + \Sigma_1 }{\left(m_\Delta +\Sigma_1\right)^2
- p^2 \left(1-\Sigma_{6}\right)^2},\\
\label{S1}
S_6&=&\frac{ 1-\Sigma_6}{\left( m_\Delta +\Sigma_1\right)^2 - p^2
\left( 1-\Sigma_{6}\right)^2}\,. \label{d6}
\end{eqnarray}
   As was to be expected, both scalar functions $S_1$ and $S_6$ have the
same poles.

\begin{figure}
\epsfig{file=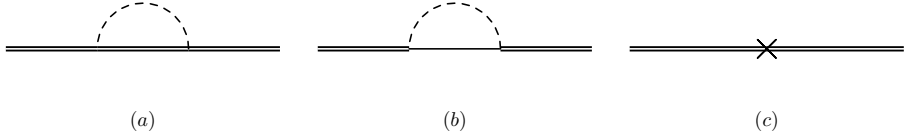, width=12truecm}
\caption[]{\label{mdeltaloops:fig} One-loop contributions to the
$\Delta$ self-energy at ${\cal O}(p^3)$.}
\end{figure}

   The pole is found by solving the equation
\begin{equation}
x=f(x)
\label{polepositioneq}
\end{equation}
with
\begin{equation}
\label{fx}
f(x)=\frac{[m_\Delta+\Sigma_1(x)]^2}{[1-\Sigma_6(x)]^2}.
\end{equation}
   Performing a loop expansion for both the function $f(x)$ as well as the
solution $y$ to Eq.\ (\ref{polepositioneq}),
\begin{eqnarray*}
f(x)&=&\sum_{i=0}^\infty \hbar^i\,f^{(i)}(x),\\
y&=&\sum_{j=0}^\infty \hbar^j y^{(j)},
\end{eqnarray*}
we obtain up to and including order $\hbar$:
\begin{displaymath}
y^{(0)}+\hbar y^{(1)}+O(\hbar^2)=
y^{(0)}+\hbar \frac{f^{(1)}(y^{(0)})}{1-f^{(0)'}(y^{(0)})}+O(\hbar^2).
\end{displaymath}
   In fact, using suitable field redefinitions
\cite{Kamefuchi:1961sb,Scherer:1994wi}, in a first step any dependence on
$p^2$ of the tree-level contribution to the self-energy can be removed,
i.e.~$f^{(0)'}(x)=0$.
   We then obtain, setting $\hbar=1$,
\begin{equation}
s=s^{(0)}+s^{(1)}+\cdots
\label{pps6eq0}
\end{equation}
for the pole $s$ of the $\Delta$ propagator to one-loop order,
where
\begin{eqnarray*}
s^{(0)}&=&\frac{\left(m_\Delta + \Sigma_1^{(0)}\right)^2}{
\left(1-\Sigma_6^{(0)}\right)^2},\\
s^{(1)}&=&
2 \frac{m_\Delta + \Sigma_1^{(0)}}{\left(1-\Sigma_6^{(0)}\right)^3}
\left[(m_\Delta+\Sigma_1^{(0)})\Sigma_6^{(1)}(p^2)
+(1-\Sigma_6^{(0)})\Sigma_1^{(1)}(p^2)\right]_{p^2=s^{(0)}}.
\end{eqnarray*}
   In a second step, using again a suitable field redefinition, the
tree-level contribution proportional to ${\cal P}_6^{\mu\nu}=
g^{\mu\nu} p\hspace{-.45em}/\,$, i.e.~$\Sigma_6^{(0)}$, can also be removed
and the result simplifies even further,
\begin{equation}
s=\left(m_\Delta + \Sigma_1^{(0)}\right)^2
+ 2 \left(m_\Delta +
\Sigma_1^{(0)}\right)
\left[\left(m_\Delta+\Sigma_1^{(0)}\right)\Sigma_6^{(1)}(p^2)
+\Sigma_1^{(1)}(p^2)\right]_{p^2=\left(m_\Delta + \Sigma_1^{(0)}\right)^2}
+\cdots.
\label{pp}
\end{equation}
Substituting $\Sigma_1^{(0)}=-4 c^\Delta_1\,M^2$ in
Eq.~(\ref{pp}) we obtain to one-loop order at ${\cal O}(p^3)$
\begin{equation}
s = m_\Delta^2 - 8\,m_\Delta\,c^\Delta_1 M^2+s_{\rm 1\, loop},
\label{spole}
\end{equation}
where the contributions to $s_{\rm 1\, loop}$ of ${\cal O}(p^3)$
result from the loop diagrams of Fig.~\ref{mdeltaloops:fig}.

   The unrenormalized contribution of Fig.~\ref{mdeltaloops:fig}~(a)
with an internal delta line is given by
\begin{eqnarray}
\Sigma^{\mu\nu}_{{\rm loop},\Delta} &= & - \frac{5 g_1^2}{3 F^2}
\Biggl\{ \frac{2}{3}I_{m_\Delta}^{\mu\nu}(411) \left(2 m_\Delta +
3\, p\hspace{-.45em}/\hspace{.1em}
-\frac{p^2 p\hspace{-.45em}/\hspace{.1em}}{m_\Delta^2} \right)
-g^{\mu\nu}I_{m_\Delta}^{\alpha\beta}(411)
\left[ g_{\alpha\beta}\left( m_\Delta +
p\hspace{-.45em}/\hspace{.1em} \right)-2 \gamma_\alpha p_\beta
\right]  \nonumber\\
&&+g^{\mu\nu} I_{m_\Delta}^{\alpha\beta\lambda}(411)
g_{\alpha\beta}\gamma_\lambda
-\frac{2}{3} I_{m_\Delta}^{\mu\nu\lambda}(411)
\frac{{\gamma_\lambda}\left({m_\Delta}^2 - p^2 \right)
+ 2 {p_\lambda}\left( {m_\Delta} +
p\hspace{-.45em}/\hspace{.1em} \right)}{m_\Delta^2}\Biggr\}.
\label{Dse1lD}
\end{eqnarray}
   The corresponding contribution to the pole reads
\begin{eqnarray}
s_{{\rm loop},\Delta} &=& \frac{5 g_1^2  m_\Delta^2}{10368 \pi^2 F^2}
\left[ 40 \Lambda^2 + 321 {{m_\Delta}}^2 -
264 {{m_\Delta}}^2 \ln\left(\frac{\Lambda}{{m_\Delta}}\right)\right]
\nonumber \\
&&+ \frac{5 g_1^2 M^2}{864 {\pi }^2 F^2} \left[
9 {m_\Delta^2} - 20 \Lambda^2 m_\Delta  {c_1^\Delta} -
151 {{m_\Delta}}^3 {c_1^\Delta} - 4 {m_\Delta^2} \left(5 -
22 {{m_\Delta}} {c_1^\Delta}\right)\ln\left(
\frac{\Lambda}{{m_\Delta}}\right)
\right] \nonumber \\
&&+\frac{25 g_1^2}{432\pi F^2} M^3 m_\Delta.
\label{polea}
\end{eqnarray}
   The unrenormalized one-loop contribution of Fig.~\ref{mdeltaloops:fig}~(b)
with an internal nucleon line is given by
\begin{equation}
\Sigma^{\mu\nu}_{{\rm loop},N}= -\frac{g^2}{F^2}\left[
\left(p\hspace{-.45em}/\hspace{.1em}+m\right) I^{\mu\nu}_m(411) +
\gamma_\lambda  I^{\lambda\mu\nu}_m(411)\right]. \label{diagrA}
\end{equation}
   The corresponding contribution to the pole reads
\begin{eqnarray}
s_{{\rm loop},N}&=& \frac{g^2}{F^2}\Biggl\{ -\frac{35
m_\Delta^4}{768\pi^2} - \frac{5 m_\Delta^2 \Lambda^2}{288
\pi^2} + \frac{5 m_\Delta^4}{96 \pi^2} \ln
\left(\frac{\Lambda}{m_\Delta}\right)\nonumber
\\ &&+ \frac{\delta m_\Delta}{96 \pi^2}
\left[ 20 m_\Delta^2 + \Lambda^2 - 20 m_\Delta^2
\ln\left(\frac{\Lambda}{m_\Delta}\right)\right] \nonumber
\\ &&+ \frac{1}{576\pi^2}  \left[ -240\delta^2 m_\Delta^2
- 15 m_\Delta^3 \Sigma_1^{(0)}  - 14 m_\Delta  \Sigma_1^{(0)} \Lambda^2
+ 60 m_\Delta^2 \left( 2\delta^2 + M^2 \right)
\ln\left(\frac{\Lambda}{m_\Delta}\right)\right]
\nonumber \\
&&- \frac{\delta}{96 \pi^2} \left[-\Sigma_1^{(0)}\Lambda^2
+  6 m_\Delta(\delta^2+M^2) \ln\left(\frac{\Lambda}{m_\Delta}\right)
+20 m_\Delta^2 \Sigma_1^{(0)}
\ln\left(\frac{\Lambda}{m_\Delta}\right) \right] \nonumber \\
&&- \frac{m_\Delta}{288\pi^2} \Biggl[ -95 \delta^3
+3 \delta M^2 + 48 \left( \delta^2 - M^2\right)^\frac{3}{2}
\ln\left(\frac{\delta - \sqrt{\delta^2-M^2}}{M}\right)
\nonumber \\
&&- 48 \delta^3
\ln\left(\frac{M}{m_\Delta}\right) + 72\delta M^2
\ln\left(\frac{M}{m_\Delta}\right) \Biggr] \nonumber \\
&&- i\frac{ m_\Delta
\left(\delta^2 - M^2\right)^\frac{3}{2}}{6\pi}\Biggr\},
\label{polepositions60eq0}
\end{eqnarray}
where
$\Sigma_1^{(0)}=-4 c_1^\Delta \,M^2$.

   After taking  the contribution of the counterterm diagram of
Fig.~\ref{mdeltaloops:fig}~(c) into account, we finally obtain
for the pole
\begin{equation}
s=m_{\Delta}^2 - 8  m_{\Delta}\, c_1^\Delta M^2
+ 2 m _\Delta  L_\Delta+ 2 m_\Delta L_N - i\,\frac{g^2}{F^2}
\frac{ m_{\Delta} \left(\delta^2 - M^2\right)^\frac{3}{2}}{6\pi}
+{\cal O}(p^4),
\label{poleren}
\end{equation}
where
\begin{eqnarray*}
L_\Delta&=&-\frac{g^2}{576\,\pi^2 F^2}\,\Biggl[ -95 \,
\delta^3 +3\,\delta\, M^2\,
+ 48\,\left( \delta^2 - M^2\right)^\frac{3}{2}
\ln\left(\frac{\delta - \sqrt{\delta^2-M^2}}{M}\right)\\
&&
+24\,\delta \left( 3\,M^2-2\,\delta^2\right)
\ln\left(\frac{M}{m_{\Delta}}\right) \Biggr],\\
L_N&=&\frac{25\,g_1^2}{864\,\pi\,F^2} \,M^3.
\end{eqnarray*}
The contribution of the renormalized loop diagrams is indeed of
order $p^3$ as suggested by the power counting.
   The so-called pole mass is given by
\begin{equation}
{m_\Delta}_{\rm pole}=m_\Delta-4 c_1^\Delta M^2+L_\Delta+L_N+{\cal O}(p^4).
\end{equation}
   Using the numerical values of Eq.\ (\ref{numericalvalues})
and the SU(6) estimate $g_1=9\, g_A/5$, we obtain
$L_\Delta=102.8$ MeV and $L_N=15.6$ MeV.
   We made use of 100 MeV \cite{Eidelman:2004wy}
as the value for $(-2)$ times the imaginary part of the pole to
fix $g=1.127$.
   Unfortunately we do not have a reliable estimate for the parameter
$c_1^\Delta$.
   For the moment, we assume it to be the same as $\tilde c_1$ of the
$\pi N$ Lagrangian and we obtain an estimate of 126 MeV to 177 MeV
for $-4 c_1^\Delta M^2$, depending on how $\tilde c_1$ is fitted
to data.

\section{Summary}

   We have considered the explicit inclusion of the $\Delta(1232)$ resonance
in baryon chiral perturbation theory.
   The requirement of the consistency of the corresponding effective field
theory in the sense of having the right number of degrees of
freedom, leads to non-trivial constraints among coupling constants
of various interaction terms. These constraints are compatible
with the symmetries underlying the effective theory.
   Implementing them in the effective Lagrangian and using the extended on-mass-shell
renormalization scheme (or the reformulated version of the IR
renormalization) in combination with the higher-derivative
formulation we obtain a consistent effective field theory with a
systematic power counting. Thus, we are in a position to calculate
low-energy processes involving pions, nucleons, and deltas to any
specified order in a small parameter expansion.
   As applications we have considered the ${\cal O}\left(p^3\right)$
contributions to the nucleon mass, the pion-nucleon sigma term,
and the pole of the $\Delta$ resonance.


\acknowledgments

The work of J.~G.~has been supported by the Deutsche
Forschungsgemeinschaft (SCHE 459/2-1).

\begin{appendix}
\section{Isospin projections}
\label{isospin_projections}
   For the isospin components we make use of the following conventions.
   Let $X=Y\otimes Z$ denote the direct product of isospin-1 and isospin-1/2
spaces, respectively.
   We parameterize a general vector $|\Psi\rangle\in X$ as
\begin{displaymath}
|\Psi\rangle=\sum_{i=1}^3\sum_{\alpha=-\frac{1}{2}}^{\frac{1}{2}}
\Psi_{i,\alpha}|i\rangle\otimes|\frac{1}{2},\alpha\rangle
=\sum_{m=-1}^1\sum_{\alpha=-\frac{1}{2}}^{\frac{1}{2}}
(-)^m \Psi^{(1)}_{-m,\alpha}|1,m\rangle\otimes |\frac{1}{2},\alpha\rangle,
\end{displaymath}
   where the two decompositions refer to a Cartesian and a spherical basis
of $Y$, respectively.
   Using the Clebsch-Gordan decomposition
$X=X_\frac{3}{2}\oplus X_\frac{1}{2}$, we describe a general isospin-3/2 state
as
\begin{displaymath}
X_\frac{3}{2}\ni |\Delta\rangle=\sum_{M=-\frac{3}{2}}^\frac{3}{2}
\Delta_M|\frac{3}{2}, M\rangle.
\end{displaymath}
   The scalar product $(\langle 1,m|\otimes \langle \frac{1}{2},\alpha|)|\Delta
\rangle$ generates the component $(-)^m \Psi_{-m,\alpha}^{(1)} $ of
the state $|\Delta\rangle$ in terms of the Clebsch-Gordan coefficient
$\langle 1,m;\frac{1}{2},\alpha|(1,\frac{1}{2}) \frac{3}{2},M\rangle$ and
the components $\Delta_M$.

   Re-expressing the spherical components in terms of Cartesian components,
we then obtain, in terms of the projection operator
$\xi^\frac{3}{2}$ of Eq.\ (\ref{xi32}),
\begin{eqnarray}
\xi^{\frac{3}{2}}_{1j}\Psi_{\mu,j} &=&
\frac{1}{\sqrt{2}}\,\left(\begin{array}{c}
\frac{1}{\sqrt{3}}\,\Delta^0_\mu-\Delta^{++}_\mu\\
\Delta^{-}_\mu-\frac{1}{\sqrt{3}}\,\Delta^+_\mu\end{array}\right),
\nonumber \\
\xi^{\frac{3}{2}}_{2j}\Psi_{\mu,j} &=&
-\frac{i}{\sqrt{2}}\,\left(\begin{array}{c}
 \frac{1}{\sqrt{3}}\,\Delta^0_\mu+\Delta^{++}_\mu\\
\Delta^{-}_\mu+\frac{1}{\sqrt{3}}\,\Delta^+_\mu\end{array}\right),
\nonumber \\
\xi^{\frac{3}{2}}_{3j}\Psi_{\mu,j}
&=& \sqrt{\frac{2}{3}}\left(\begin{array}{c}\Delta^{+}_\mu\\
\Delta^{0}_\mu\end{array}\right). \label{phfields}
\end{eqnarray}
   This phase convention agrees with Ref.\ \cite{Tang:1996sq}
but is opposite to Ref.\ \cite{Hemmert:1997ye}.

\section{Decomposition}
\label{decomposition}
   For the decomposition of the dressed propagator and the self-energy of the
$\Delta$ resonance of Eqs.~(\ref{deltase}) and (\ref{deltapr}) we
make use of the basis
\begin{eqnarray}
&&{\cal P}_1^{\mu\nu}=g^{\mu\nu},\quad
{\cal P}_2^{\mu\nu} = \gamma^{\mu}\gamma^{\nu},\quad
{\cal P}_3^{\mu\nu} = p^{\mu}\gamma^{\nu}, \quad
{\cal P}_4^{\mu\nu} = \gamma^{\mu}p^{\nu},\quad
{\cal P}_5^{\mu\nu} = p^{\mu}p^{\nu},
\nonumber\\
&&
{\cal P}_6^{\mu\nu} = p\hspace{-.45em}/\hspace{.1em} g^{\mu\nu},\quad
{\cal P}_7^{\mu\nu} = p\hspace{-.45em}/\hspace{.1em} \gamma^{\mu}\gamma^{\nu},
\quad
{\cal P}_8^{\mu\nu} = p\hspace{-.45em}/\hspace{.1em} p^{\mu}\gamma^{\nu},\quad
{\cal P}_9^{\mu\nu} = p\hspace{-.45em}/\hspace{.1em} \gamma^{\mu}p^{\nu},\quad
{\cal P}_{10}^{\mu\nu}=p\hspace{-.45em}/\hspace{.1em} p^\mu p^\nu.
\label{basis}
\end{eqnarray}

\section{Loop integrals}
\label{loop_integrals}
   The loop integrals of Eq.\ (\ref{intdefIabcd}) have been
calculated using the method of dimensional counting
\cite{Gegelia:1994zz}:

\begin{equation}
I^{\mu\nu}_m(411)= g^{\mu\nu} A + \frac{p^\mu p^\nu}{p^2}B,
\label{2tensint}
\end{equation}
where
\begin{eqnarray*}
A &=&
\frac{1}{192 p^2 \pi^2}\,\Biggl[m^4-2 M^2 m^2+3 p^2 m^2+M^4+3
M^2 p^2+p^2 \Lambda ^2 \nonumber \\
&&+ 2 M^2 \left(-m^2+M^2+p^2\right) \ln\left(\frac{M}{m}\right)+2 p^2
\left(-3 m^2-3 M^2+p^2\right) \ln\left(\frac{\Lambda}{m}\right)\Biggr]
\nonumber \\
&&-\frac{I_f}{12 p^2}\left[ m^4-2 \left(M^2+p^2\right)
m^2+\left(M^2-p^2\right)^2\right],
\nonumber \\
B&=&- \frac{1}{96 p^2 \pi ^2} \Biggl[2 m^4-4 M^2 m^2-5 p^2 m^2+2
M^4-3 (p^2)^2+M^2 p^2 \nonumber \\
&&+ 4 M^2 \left(-m^2+M^2+p^2\right) \ln\left(\frac{M}{m}\right)
+4 (p^2)^2 \ln\left(\frac{\Lambda }{m}\right)\Biggr]\nonumber\\
&&+ \frac{I_f}{3 p^2} \left[m^4-2
\left(M^2+p^2\right) m^2+M^4+(p^2)^2+M^2 p^2\right].
\end{eqnarray*}
\begin{equation}
I^{\mu\nu\rho}_m(411)= \frac{g^{\mu \nu}p^\rho+g^{\mu \rho}
p^\nu + g^{\nu \rho}p^\mu}{p^2}C
+\frac{p^\mu p^\nu p^\rho}{(p^2)^2} D,
\label{3tensint}
\end{equation}
where
\begin{eqnarray*}
C&=&\frac{1}{4608 p^2 \pi^2}
\Biggl\{ 12 m^6-36 M^2 m^4-42 p^2 m^4 + 36 M^4 m^2 - 60 (p^2)^2 m^2
+24 M^2 p^2 m^2\nonumber\\
&&-12 M^6+3 (p^2)^3 -36 M^2 (p^2)^2  + 18 M^4 p^2
- 8 (p^2)^2  \Lambda ^2\nonumber\\
&&-24 M^2 \left[ m^4-2 \left(M^2+p^2\right)
m^2+M^4+(p^2)^2-M^2 p^2\right] \ln\left(\frac{M}{m}\right)\nonumber
\\
&&+ 24 (p^2)^2  \left(4\,m^2+2\,M^2-p^2\right)
\ln\left(\frac{\Lambda}{m}\right)\Biggr\} \nonumber \\
&&- \frac{I_f}{24 (p^2)^2} \Biggl\{ m^6-3 \left(M^2+p^2\right)
m^4+\left[3 M^4+2 p^2 M^2+3 (p^2)^2\right] m^2
   \nonumber \\
&&- \left(M^2-p^2\right)^2
   \left(M^2+p^2\right)\Biggr\}
\nonumber \\
D&=&\frac{1}{768p^2\pi^2}
\Biggl\{-12 m^6+36 M^2 m^4+42 p^2 m^4-36 M^4 m^2-52 (p^2)^2 m^2 \nonumber \\
&&- 48 M^2 p^2 m^2+12 M^6-19 (p^2)^2+4 M^2 (p^2)^2+6 M^4 p^2 \nonumber \\
&&+ 24 M^2 \left[ m^4-2 \left(M^2+p^2\right) m^2+M^4+(p^2)^2+M^2
p^2\right] \ln\left(\frac{M}{m}\right)
+24 (p^2)^3 \ln\left(\frac{\Lambda}{m}\right)\Biggr\} \nonumber \\
&&-\frac{I_f}{{4p^2}}\Biggl\{-m^6+3 \left(M^2+p^2\right)m^4
-\left[3 M^4+4 p^2 M^2+3 (p^2)^2\right] m^2\nonumber \\
&&+ M^6+(p^2)^3+M^2 (p^2)^2+M^4 p^2\Biggr\}.
\end{eqnarray*}
   In the above expressions $I_f$ is given by \cite{Fuchs:2003qc}
\begin{equation}
\label{If} I_f = \frac{1}{16\pi^2}\left[\frac{p^2-m^2+M^2}{p^2}
\ln\left(\frac{M}{m}\right) +\frac{2\,m\,M}{p^2}F(\Omega)\right],
\end{equation}
where
\begin{eqnarray*}
F(\Omega) &=& \left \{ \begin{array}{ll}
\sqrt{\Omega^2-1}\ln\left(-\Omega-\sqrt{\Omega^2-1}\right),&\Omega\leq -1,\\
\sqrt{1-\Omega^2}\arccos(-\Omega),&-1\leq\Omega\leq 1,\\
\sqrt{\Omega^2-1}\ln\left(\Omega+\sqrt{\Omega^2-1}\right)
-i\pi\sqrt{\Omega^2-1},&1\leq \Omega,
\end{array} \right.
\end{eqnarray*}
with
\begin{displaymath}
\Omega=\frac{p^2-m^2-M^2}{2mM}.
\end{displaymath}

\end{appendix}

\end{document}